\documentclass[10pt]{article}
\usepackage{graphicx}
\usepackage{color}
\usepackage{cite}
\usepackage{multirow}

\usepackage{setspace} 
\usepackage[labelfont=bf,labelsep=period,justification=raggedright]{caption}

\makeatletter
\renewcommand{\@biblabel}[1]{\quad#1.}
\makeatother

\date{}

\newcounter{figcount}
\renewcommand{\thefigcount}{\arabic{figcount}}
\newcommand{\figcount}{\refstepcounter{figcount}%
           Figure~\thefigcount:~}

\newcounter{tabcount}
\renewcommand{\thetabcount}{\arabic{tabcount}}
\newcommand{\tabcount}{\refstepcounter{tabcount}%
           Table~\thetabcount:~}

\newcommand{\FIG}{Figure~}
\newcommand{\FIGS}{Figures~}
\newcommand{\TAB}{Table~}
\newcommand{\TABS}{Tables~}

\begin{document}

\begin{flushleft}
{\Large \textbf{Suicide ideation of individuals in online social networks}
}
\\
Naoki Masuda${}^{1*}$, Issei Kurahashi${}^2$, Hiroko Onari$^2$
\\
\bf{1} Department of Mathematical Informatics,
The University of Tokyo,
7-3-1 Hongo, Bunkyo, Tokyo 113-8656, Japan
\\
\bf{2} iAnalysis LLC,
2-2-15 Minamiaoyama, Minato-ku, Tokyo 107-0062, Japan
\\
$\ast$ Email: masuda@mist.i.u-tokyo.ac.jp
\end{flushleft}

\section*{Abstract}
Suicide explains the largest number of death tolls among Japanese adolescents in their twenties and thirties. Suicide is also a major cause of death for adolescents in many other countries. Although social isolation has been implicated to influence the tendency to suicidal behavior, the impact of social isolation on suicide in the context of explicit social networks of individuals is scarcely explored. To address this question, we examined a large data set obtained from a social networking service dominant in Japan. The social network is composed of a set of friendship ties between pairs of users created by mutual endorsement. We carried out the logistic regression to identify users' characteristics, both related and unrelated to social networks, which contribute to suicide ideation. We defined suicide ideation of a user as the membership to at least one active user-defined community related to suicide. We found that the number of communities to which a user belongs to, the intransitivity (i.e., paucity of triangles including the user), and the fraction of suicidal neighbors in the social network, contributed the most to suicide ideation in this order. Other characteristics including the age and gender contributed little to suicide ideation. We also found qualitatively the same results for depressive symptoms.


\section*{Introduction}

Suicide is a major cause of death in many countries.
Japan possesses the highest suicide rate among the OECD countries
in 2009 \cite{Chambers2010Guardian}.
In fact, suicide explains the largest number of death cases
for Japanese adolescents in their twenties and thirties
\cite{Chambers2010Guardian}.
Suicide is also a major cause of death for youths in other
countries including the United States \cite{StatAbstUSA2012}.

Since the seminal sociological study by Durkheim in the late nineteenth
century \cite{Durkheim1951book}, suicides have been studied for both
sociology interests and public health reasons.  In
particular, Durkheim and later scholars pointed out that social isolation,
also referred to as the lack of social integration, is a significant
contributor to suicidal behavior
\cite{Durkheim1951book,Trout1980SLTB,Joiner2005ARP,Wray2011ARS}.
Roles of social isolation in inducing other physical and
mental illnesses have also been examined \cite{Putnam2000book}.
Conceptual models that inherit Durkheim's idea
also claim that
social networks affect general health conditions including tendency to
suicide \cite{Pescosolido1989ASR,Bearman1991SF,Berkman2000SSM,Kawachi2001JUH}.

Social network analysis provides a pragmatic method to quantify social
isolation
\cite{Wasserman1994book,Newman2010book}.  In their seminal work,
Bearman and Moody explicitly studied the relationship between suicidal
behavior and egocentric social networks for American adolescents using
data obtained from a national survey (National Longitudinal Study of
Adolescent Health) \cite{Bearman2004AJPH}.  They showed that, among
many independent variables including those unrelated to social
networks, a small number of friends and a small fraction of triangles
to which an individual belongs significantly contribute to suicide
ideation and attempts.  A small number of friends is an intuitive
indicator of social isolation.  Another study derived from
self reports from Chinese adolescents also supports this idea in a
quantitative manner \cite{Cui2010CCHD}.  The paucity of triangles, or
intransitivity \cite{Wasserman1994book}, also
characterizes social isolation \cite{Bearman2004AJPH}.
Individuals without triangles are considered to
lack membership to social groups even if they have many friends
\cite{Krackhardt1999chapter}; social groups are often approximated by
overlapping triangles \cite{Palla2005Nature,Onnela2007PNAS}.

Nevertheless, the structure of the Bearman--Moody study \cite{Bearman2004AJPH}
implies that our understanding of relationships between social
networks and suicide is still limited.
First, in the survey, a respondent was allowed to list best five
friends of each gender. However, many respondents would generally have
more friends. The imposed upper limit may distort
network-related personal quantities such as the number of friends and
triangles.  Second, their study was confined inside each school in the
sense that only in-school names are matched. If a respondent X
named two out-school friends that were actually friends of each other,
the triangle composed of these three individuals was dismissed
from the analysis.  Therefore, the accuracy of the
triangle counts in their study may be limited such that the
relationship between intransitivity and suicidal
behavior remains elusive.

In the present study, we examine the relationship between social
networks
and suicide ideation using a data set obtained from a
dominant social networking service (SNS) in Japan, named mixi.
Our approach addresses limitations
in the previous study \cite{Bearman2004AJPH}.  First, an entire social
network of users is available, where a link between two users
represents explicit bidirectional friendship endorsed by both users. Some users have quite a large number of friends,
as in general social networks \cite{Newman2010book}.
Second, for the same reason, we can accurately calculate
the number of triangles for each
user. An additional feature of 
the present data set is that the sample is
relatively diverse because anybody can register for free.
In contrast, the respondents were
7 to 12 graders in schools in the Bearman--Moody study.

A function of mixi relevant to this study is
user-defined communities. A community is a group of users
that get together under a common interest, such as hobby, affiliation,
or creed. A user-defined community
of mixi is often composed of users that have not known
each other beforehand.
Although some SNSs have user-defined communities, and their dynamics
were studied \cite{Backstrom2006SIGKDD}, major SNSs including Facebook
do not own this type of user-defined communities.
We define suicide ideation by the membership of a user to at least one
community related to suicide. Then, we statistically compare 
users with and without suicide ideation in terms of
users' properties including
those related to egocentric networks.

\section*{Results}\label{sec:results}

\subsection*{Multivariate logistic regression}

We defined the group of users with suicide ideation and the control group of users, as described in Methods.
Table~\ref{tab:independent vars suicide} indicates that
the difference in the mean of each independent variable (see Methods for the definition of the independent variables) between the
suicide and control groups is significant
($p<0.001$, Student's $t$-test).
We also verified that
the distributions of each independent variable are also significantly
different between the two groups ($p<0.0001$, Kolmogorov-Smirnov test).

The results obtained from the multivariate logistic regression are summarized in
\TAB\ref{tab:multivariate logistic}.
The VIF values (see Methods)
are much less than 5 for all the independent
variables. The three types of correlation coefficients between pairs
of the independent variables are also sufficiently small
(\TAB\ref{tab:correlation coefficient}). On these bases, we justify
the application of the multivariate logistic regression to our data.

The odds ratio (OR)
values shown in Table~\ref{tab:multivariate logistic} suggest
the following. A one-year older user is
1.00463 times more likely to belong to the suicide group than the control group
on average.
Likewise, being female,
membership to one community,
having one friend, an increase in $C_i$ by 0.01,
an increase in the fraction of friends in the suicide group (i.e., homophily variable) by 0.01,
and one day of the registration period
make a user 0.821,
1.00733, 0.99790, $0.0093^{0.01}=0.95$,
$\left(2.22\times 10^{12}\right)^{0.01}=1.33$, and
0.999383 times
more likely to belong to the suicide group, respectively.
For all the independent variables, the 95
\% confidence intervals of the ORs do not contain unity, and
the $p$-values are small. Therefore,
all the independent variables significantly contribute
to the regression. In addition, because the AUC (see Methods)
is large (i.e. 0.873),
the estimated multivariate logistic model captures
much of the variation in the user's behavior, i.e., whether to belong
to the suicide group or not.

\subsection*{Univariate logistic regression}

All the independent variables significantly
contribute to the multivariate regression probably because of
the large sample size of our data set. Therefore, we
carried out the univariate logistic regression between the dependent
variable (i.e., membership to the suicide versus control group) and
each independent variable to better clarify the contribution of each
independent variable.

The results obtained from the univariate logistic regression
are shown in \TAB\ref{tab:univariate logistic}.
Although the $p$-value for each independent variable is small,
the AUC value considerably varies between different independent variables.
The ORs for the
community number, local
clustering coefficient, homophily,
and registration period are consistent between the multivariate and
univariate regressions. For example, both regressions indicate that
a user with a large community number tends to belong to the suicide group.
These independent variables also yield large
AUC values under the univariate regression.

The community number makes
by far the largest contribution among the seven independent variables. The 
AUC value obtained from the univariate regression (0.867)
is close to that obtained by the
multivariate regression (0.873).

The independent variable with the second largest explanatory power
is the local clustering coefficient (AUC
$=$ 0.690).
The results are consistent with the previous ones \cite{Bearman2004AJPH}.
We stress that we reach this conclusion
using a data set whose full social network is available.

The homophily variable makes the third largest contribution
(AUC $=$ 0.643).
Although we refer to this independent variable as
homophily (see Methods),
the effect of this variable is in fact interpreted as either
homophily or contagion \cite{Aral2009PNAS,Shalizi2011SMR}.
Nevertheless, the result is consistent with previous
claims that
suicide is contagious (for recent accounts, see
\cite{Mann2002AIM,Baller2002ASR,Romer2006JComm,Hedstrom2008SocForces,Baller2009JHSB,Wray2011ARS}; but see \cite{Gould1989SLTB} for a critical review) and that
other related states such as depressive symptoms are contagious
\cite{Christakis2009book,Rosenquist2011MolPsyc} (but see
\cite{Lyons2011SPP,VanderWeele2011SMR}).

The effect of the age, gender, and degree (i.e., number of friends), on suicide
ideation is small, yielding small AUC values, close to the minimum
value $0.5$ (\TAB\ref{tab:univariate logistic}).  In addition, the ORs
for these variables are inconsistent between the multivariate and
univariate regressions. For example, a female user is more likely to
belong to the suicide group according to the univariate regression
and vice versa according to the multivariate regression.  Therefore,
we conclude that these three independent variables do not
explain suicide ideation.

The registration period also yields 
a small AUC value (i.e., 0.545).
Therefore, suicide ideation depends on
the community number, local clustering coefficient, and homophily variable
not because they commonly depend on the registration period.

\subsection*{Depressive symptoms}

Our data set allows us to investigate correlates between users' other
characteristics and the independent variables if the
characteristics have corresponding used-defined communities
in the SNS. We repeated the
same series of analysis for depressive symptoms, which are suggested to be
implicated in suicidal behavior
\cite{Mann2002AIM,Joiner2005ARP,Brezo2006APS}.
A user is defined to own depressive symptoms when the user
belongs to at least one of the seven depression-related communities (Methods).

The statistics of the independent variables for the depression group
are compared with those for the control group
in \FIGS\ref{fig:community number}, \ref{fig:p(k)}, \ref{fig:C(k)}, and \TAB\ref{tab:independent vars depression}.
Each independent variable in the
depression and control groups is significantly different in terms of
the mean ($p<0.0001$, Student's $t$-test; see 
\TAB\ref{tab:independent vars depression})
and distribution ($p<0.0001$, Kolmogorov-Smirnov test).

We applied the multivariate and univariate logistic regressions to
identify independent variables that contribute to depressive symptoms (i.e.,
membership to the depression group).
The control group is the same as that used for the analysis of suicide ideation.
The results are shown in \TABS\ref{tab:multivariate logistic depression} and \ref{tab:univariate logistic depression}. The VIF
values shown in \TAB\ref{tab:multivariate logistic depression} and the correlation coefficient values shown
in \TAB\ref{tab:correlation coefficient} qualify the use of the multiple logistic
regression. The results are qualitatively the same as those for the
suicide case.

\section*{Discussion}

We investigated relationships between suicide ideation and personal
characteristics including social network variables
using the data obtained from a major SNS in Japan.
We found that an increase in the community number (i.e., the number of
user-defined communities to which a user belongs), decrease in the local
clustering coefficient (i.e., local density of triangles, or transitivity),
and increase in the homophily variable (i.e., fraction of neighboring users
with suicide ideation) contribute to suicide ideation by the largest
amounts in this order.
In addition, the results are qualitatively the same
when we replaced suicide ideation by depressive symptoms.
Remarkably, the most significant three variables represent online social
behavior of users rather than demographic properties
such as the age and gender.

Our result that the age and gender little influence suicide ideation
is inconsistent with
previous findings \cite{Wray2011ARS}. The weak age
effect in our result 
may be because the majority of registered users is young; the mean
age of the users in 
the control group is 27.7 years old (\TAB\ref{tab:independent
vars suicide}). Nevertheless, we stress that suicide is a problem
particularly among young generations to which a majority of the users belong.

We concluded that the node degree little explains
suicide ideation. In contrast, previous studies showed
that suicidal behavior is less observed for individuals with 
more friends \cite{Bearman2004AJPH,Cui2010CCHD}.
It has also been a
long-standing claim that social isolation elicits suicidal behavior
\cite{Durkheim1951book,Trout1980SLTB,Joiner2005ARP,Wray2011ARS}.
As compared to typical users, some users
may spend a lot of time online to gain many ties with other users and 
belong to many communities on the SNS. Such a user may be active exclusively online and feel lonely, for example, to be prone to
suicide ideation. Although this is a mere conjecture,
such a mechanism would also explain the strong contribution of the community number to suicide ideation revealed in our analysis.
In contrast, many people nowadays, especially the young, regularly
devote much time to online activities including SNSs
\cite{Martin2010Nielsen}.
Therefore, the data obtained from SNSs may
capture a significant part of users' real lives.

Because
mixi enjoys a large number of users
and implements the user-defined community as a main function,
its user-defined communities cover virtually all major topics.
Therefore, 
applying the present methods to other psychiatric illness and symptoms, such as
schizophrenia, bipolar disorder, and alcohol abuse, as well as positive symptoms
may be profitable.

Our studies are limited in some aspects.  First, we identified suicide
ideation with the membership to a relevant community, but not with
suicide attempts or committed suicides.  Second, membershipship to a
relevant community may not even imply suicide ideation. Users
may enter the suicide group because they have encountered suicide
among their friends or family.  Third, our data are a specific sample
of individuals from a general population. This criticism applies to
any work that relies on SNS data. However, it is particularly
pertinent when one focuses on individuals' chracteristics (e.g.,
personality and attitudes) rather than collective phenomena online
(e.g., contagion on SNSs). Although it is beyond the scope of the
current study, quantifying the 
extent to which our sample accurately represents
general populations remains a future challenge.

\section*{Methods}

\subsection*{Data}\label{sub:data}

Mixi is a major SNS in Japan.  It started to operate on March 2004
and enjoys more than $2.7 \times 10^7$
registered users as of March 2012.
Similar to other known
SNSs, users of mixi can participate in
various activities such as making friendship
with other users, writing microblogs, sending instant
messages to others, uploading photos, and playing online games. Registration is
free.
See \cite{Yuta2007arxiv} for a previous study of the mixi social network.

In mixi, there were more than $4.5\times 10^6$
user-defined communities on various topics as of April 2012.
Users can join a user-defined 
community if the owner personally permits or the owner allows anybody to join it.

We identified suicide ideation with the membership of a user to at least one suicidal community. To define suicidal community, which is sufficiently
active, we first
selected communities satisfying the following five criteria:
(1) The name included the word
``suicide'' (``jisatsu'' in Japanese),
(2) there were at least 1000 members 
on November 2, 2011, (3)
there were at least 100 comments posted on October, 2011, which were directed to other comments or topics,
(4) there were at least three independent topics on which
comments were made on October, 2011,
and (5) the condition for admission was made open to public.
Seven communities met these criteria. Then, we excluded one
community whose name indicated that it concentrated on
methodologies of committing suicide and two communities
whose names indicated that they encouraged members to
live with hopes (one contained the word ``want to live'', and the other contained the word ``have a fun'' in their names; translations by the authors).

As a result, four communities 
were qualified as suicidal communities.
The user statistics of these communities
are shown in \TAB\ref{tab:suicide comms}.
A user that belongs to at least one suicidal community is defined to possess suicide ideation.
To exclude inactive users, we restricted
ourselves to the set of active users.
The active user was defined as users that existed as of January 23, 2012 and 
logged on to mixi in more than
20 days per month on average from August through December 2011. A similar definition was used in a previous study of the Facebook social network \cite{Ugander2011arxiv}.
We also discarded users with zero or one friend
on mixi because the triangle count described below
was undefined for such users. Despite this exclusion, the remaining data allowed us to examine the effect of social isolation in terms of the 
degree, i.e., number of neighbors, because the degree was widely distributed between 2 and 1000.
There were 9990 active users with suicide ideation (suicide group).

We statistically compared the users in the 
suicide group with users without suicide ideation.
Because the number of users was huge,
we randomly selected 228949 active users that possessed at least two friends and
belonged to neither of the seven candidates of the suicidal community defined above nor the ten candidates of the depression-related community defined below.
We call this set of users the control group.

The employees of mixi deleted private information irrelevant to the present study and encrypted the relevant private information
before we analyzed the data.
In addition, we conducted all the analysis
in the central office of mixi located in Tokyo using
a computer that was not connected to Internet.

\subsection*{Statistical models}\label{sub:statistical methods}

The dependent variable that represents the level of suicide ideation
is binary, i.e., whether a user belongs to a suicidal community or
not. Therefore, we used univariate and multivariate logistic
regressions. To check the multicollinearity between independent variables to
justify the use of the multivariate logistic regression, we
carried out two subsidiary analysis. First, we measured the variance
inflation factor (VIF) for each independent variable (see
\cite{Stine1995AmStat,Tuffery2011book} and references therein). The VIF is
the reciprocal of the fraction of the variance of the
independent variable that is not explained by linear combinations
of the other independent variables. It is recommended that the VIF
value for each independent variable is smaller than 10 (preferably smaller than 5)
for the multivariate logistic regression to be valid. Second, we
measured the Pearson, Spearman, and Kendall correlation coefficients
between the independent variables.

To quantify the explanatory power of the logistic model, we measured
the area under the receiver operating characteristic curve (AUC) for
each fit (e.g., \cite{Tuffery2011book}). The receiver operating
characteristic curve is the trajectory of the false positive (i.e.,
fraction of users in the control group that are mistakenly classified
into the suicide group on the basis of the linear
combination of the independent variables) and the true positive (i.e.,
fraction of users in the suicide group correctly classified into the
suicide group), when the threshold for classification is varied.  The
AUC value falls between 0.5 and 1. A large AUC value indicates
that the logistic regression fits well to the data in the sense that
users are accurately classified into suicide and control groups.

\subsection*{Independent Variables}

We considered seven independent variables.  Their univariate
statistics for the suicide and control groups are shown in
\TAB\ref{tab:independent vars suicide}.

\textit{Demographics}. Demographic independent variables include age
and gender.  Our analysis does not include ethnic components because
most users are Japanese-speaking Japanese; mixi provides
services in Japanese.  Other demographic, socioeconomic, and personal
characteristic variables such as residence area, occupation,
company/school, and hobby, were not
used because they were unreliable. In fact, many users leave them
blank or do not fill them consistently, probably because they do
not want to disclose them.

\textit{Community number}. The number of user-defined
communities that a user
belongs to was adopted as an independent variable. We refer to this
quantity
as community number.
The community number obeys a long tailed
distribution for both suicide and control groups
(\FIG\ref{fig:community number}). The mean
is quite different between the two groups
(\TAB\ref{tab:independent vars suicide}).

\textit{Degree}. 
When a user sends a request to another user and the recipient accepts
the request, the pair of users form an undirected social tie, called
Friends. A web of Friends defines a social network of mixi.
We adopted degree as the most basic network-related independent variable.
The degree is the number of neighbors (i.e., Friends),
and denoted by $k_i$ for user $i$.
The system of mixi allows a user to own at most degree 1000.
As is consistent with the previous analysis of a much smaller
data set of mixi \cite{Yuta2007arxiv}, the degree distributions for both
groups are long tailed (\FIG\ref{fig:p(k)}).
A small degree is an indicator of social isolation.

\textit{Local clustering coefficient}
We quantified transitivity, or the density of triangles around a user,
by the local clustering
coefficient, denoted by $C_i$ for user $i$. 
A directed-link version of the
same quantity was used in the Bearman--Moody study.
For user $i$ having degree $k_i$,
there can be maximum $k_i(k_i-1)/2$ triangles
that include user $i$. We defined $C_i$ as the actual number of
triangles that included $i$ divided by $k_i(k_i-1)/2$.
Examples are shown in \FIG\ref{fig:concept C_i}.
By definition, $0\le C_i\le 1$.
We discarded the users with $k_i\le 1$
because $C_i$ was defined only for users with $k_i\ge 2$.
$C_i$ quantifies the extent to which neighbors of user $i$ are
adjacent to each other \cite{Watts1998Nature,Newman2010book}.
If $C_i$ is large, the user is probably embedded 
in close-knit social groups
\cite{Wasserman1994book,Watts1998Nature,Newman2010book}.
A small $C_i$ value is an indicator of social isolation.
As in many networks \cite{Newman2010book}, $C_i$
decreases with $k_i$ in both suicide and control groups
(\FIG\ref{fig:C(k)}). The results are consistent with those in the previous study in which the average $C_i$ obtained without categorizing users 
is roughly proportional to $k_i^{-0.6}$ \cite{Yuta2007arxiv}.
Therefore, we carefully distinguished the
influence of $k_i$ and $C_i$ on suicide ideation by combining
univariate and multivariate regressions.

\textit{Homophily}.
Suicide may be a contagious phenomenon
(e.g.,
\cite{Mann2002AIM,Baller2002ASR,Romer2006JComm,Hedstrom2008SocForces,Baller2009JHSB,Wray2011ARS}). If
so, a user is inclined to suicide ideation when a neighbor in the social
network is.
Therefore, we adopted the fraction of
neighbors with suicide ideation as an independent variable.
It should be noted that, even if a user with suicide ideation has relatively many friends with suicide ideation, it does not necessarily imply that suicide is contagious. Homophily may be a cause of such assortativity. In this study, 
we did not attempt to distinguish the effect of imitation and homophily.
The differentiation would require
analysis of temporal data
\cite{Aral2009PNAS,Shalizi2011SMR}. Nevertheless, for a notational
reason, we refer to the fraction of neighbors as the 
homophily variable.

\textit{Registration period}. A user that registered to mixi long time
ago may be more active and own more resources in mixi
than new users. Such an experienced user may tend to simultaneously have, for example,
a large community number, large degree, and perhaps
high activities in various communities including suicidal ones.
To control for this factor,
we measured the registration period defined as
the number of days between the registration date and 
January 23, 2012.

\subsection*{Analysis of depressive symptoms}

To define depression-related community,
we identified the communities satisfying the five
criteria as in the case of suicidal community, but with the term suicide in the
community name replaced by depression (``utsu'' in Japanese).
There were ten such communities.
We excluded three of them because their names include positive
words
(let's overcome, resume one's place in society, cure; translations by the authors).
We defined the remaining seven communities, summarized in
\TAB\ref{tab:depression comms}, to represent depressive symptoms of users.
The depression group is the set of active users that belongs to at least one
depression-related community listed in \TAB\ref{tab:depression comms}.
The depression group contains 24410 users.

\subsection*{Ethics statement}

Mixi approved the provision of the data.

\section*{Acknowledgments}

We thank mixi, Inc. for providing us with their data and Taro Takaguchi for careful reading of the manuscript. 


\newpage
\clearpage

\section*{Figure captions}

\figcount\label{fig:community number}
Distribution of the community number (i.e., number
of communities to which a user belongs) for the suicide, depression, and control groups. We set 
the bin width for generating the histogram to 50.
The abrupt increase in the distribution at
1000 communities for the suicide and depression groups is owing to the
restriction that a user can belong to at most 1000 communities.

\bigskip

\noindent
\figcount\label{fig:p(k)}
Complementary cumulative distribution of the degree (i.e., fraction of
users having the degree larger than a specified value)
for the suicide, depression, and control groups.

\bigskip

\noindent
\figcount\label{fig:C(k)}
Dependence of the mean local clustering coefficient on the degree
for the suicide, depression, and control groups.
Each data point $C(k)$ for degree $k$ 
is obtained by averaging $C_i$ over the users in a group 
with degree $k$. Large fluctuations of $C(k)$
at large $k$ values are caused by
the paucity of users having large $k$. 

\bigskip

\noindent
\figcount\label{fig:concept C_i}
Examples of the degree ($k_i$) and the
local clustering coefficient ($C_i$). The shown values of $k_i$ and $C_i$ 
are for the nodes shown by the filled circles.

\newpage

\section*{Table captions}

\noindent
\tabcount\label{tab:independent vars suicide}
Univariate statistics of independent variables for the suicide and control
groups. The $p$-value for the gender is
based on the Chi-square test. The $p$-values for the other independent variables are based on the Student's $t$-test. Also shown are the statistics of two auxiliary variables that are not used in the logistic regression, i.e., the number of suicidal communities to which the user belongs and the number of days on which the user logged on to mixi. The $p$-value for the number of log-on days
is based on the Student's $t$-test.
SD: standard deviation.

\bigskip

\noindent
\tabcount\label{tab:multivariate logistic}
Multivariate logistic regression of suicide ideation on individual and network variables. OR: odds ratio; CI: 95 \% confidence interval; VIF: variance inflation factor.

\bigskip

\noindent
\tabcount\label{tab:correlation coefficient}
Correlation coefficients between pairs of independent
variables for the suicide, depression, and control groups.
P: Pearson; S: Spearman; K: Kendall correlation coefficients.

\bigskip

\noindent
\tabcount\label{tab:univariate logistic}
Univariate logistic regression of suicide ideation on individual and network variables. OR: odds ratio; CI: 95 \% confidence interval; AUC: area under the curve.

\bigskip

\noindent
\tabcount\label{tab:independent vars depression}
Univariate statistics of independent variables for the depression and control groups. The values for the control group are equal to those shown in \TAB\ref{tab:independent vars suicide} except for those of the homophily variable. The homophily is defined as the fraction of neighbors belonging to the depression group in this table, whereas it is defined as the fraction of neighbors belonging to the suicide group in \TAB\ref{tab:independent vars suicide}.
The $p$-value for the gender is
based on the Chi-square test. The $p$-values for the other variables are based on the Student's $t$-test.
SD: standard deviation.

\bigskip

\noindent
\tabcount\label{tab:multivariate logistic depression}
Multivariate logistic regression of depressive symptoms on individual and network variables. OR: odds ratio; CI: 95 \% confidence interval; VIF: variance inflation factor.

\bigskip

\noindent
\tabcount\label{tab:univariate logistic depression}
Univariate logistic regression of depressive symptoms on individual and network variables. OR: odds ratio; CI: 95 \% confidence interval; AUC: area under the curve.

\bigskip

\noindent
\tabcount\label{tab:suicide comms}
Statistics of suicidal communities.

\bigskip

\tabcount\label{tab:depression comms}
Statistics of depression-related communities.
For a technical reason, we collected the number of members 
for communities 1, 2, 3, and 6 on November 2, 2011 and communities 4, 5 and 7 on November 4, 2011.

\newpage

\begin{flushleft}
Figure 1
\end{flushleft}
\begin{center}
\includegraphics[width=10cm]{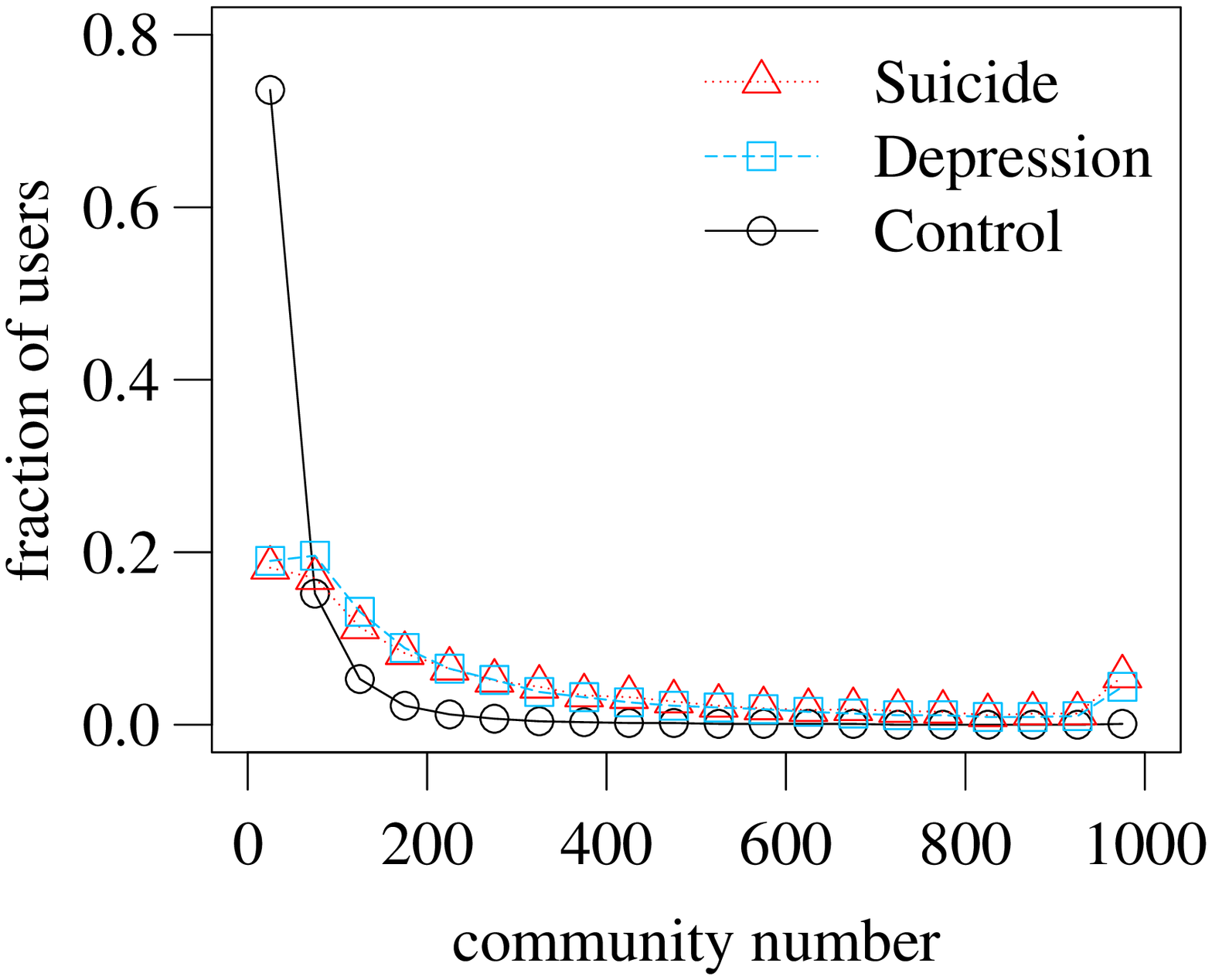}
\end{center}

\clearpage

\begin{flushleft}
Figure 2
\end{flushleft}
\begin{center}
\includegraphics[width=10cm]{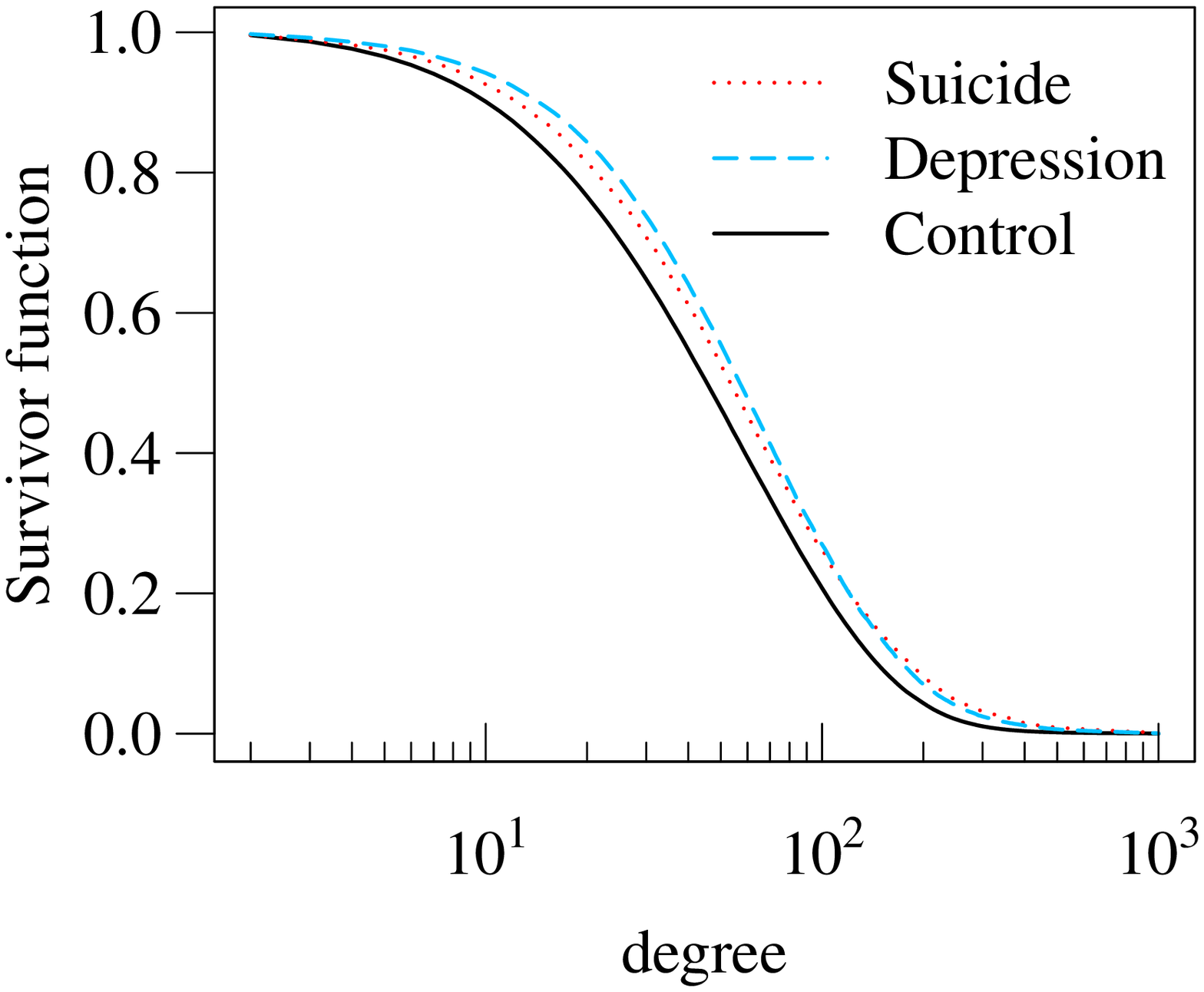}
\end{center}

\clearpage

\begin{flushleft}
Figure 3
\end{flushleft}
\begin{center}
\includegraphics[width=10cm]{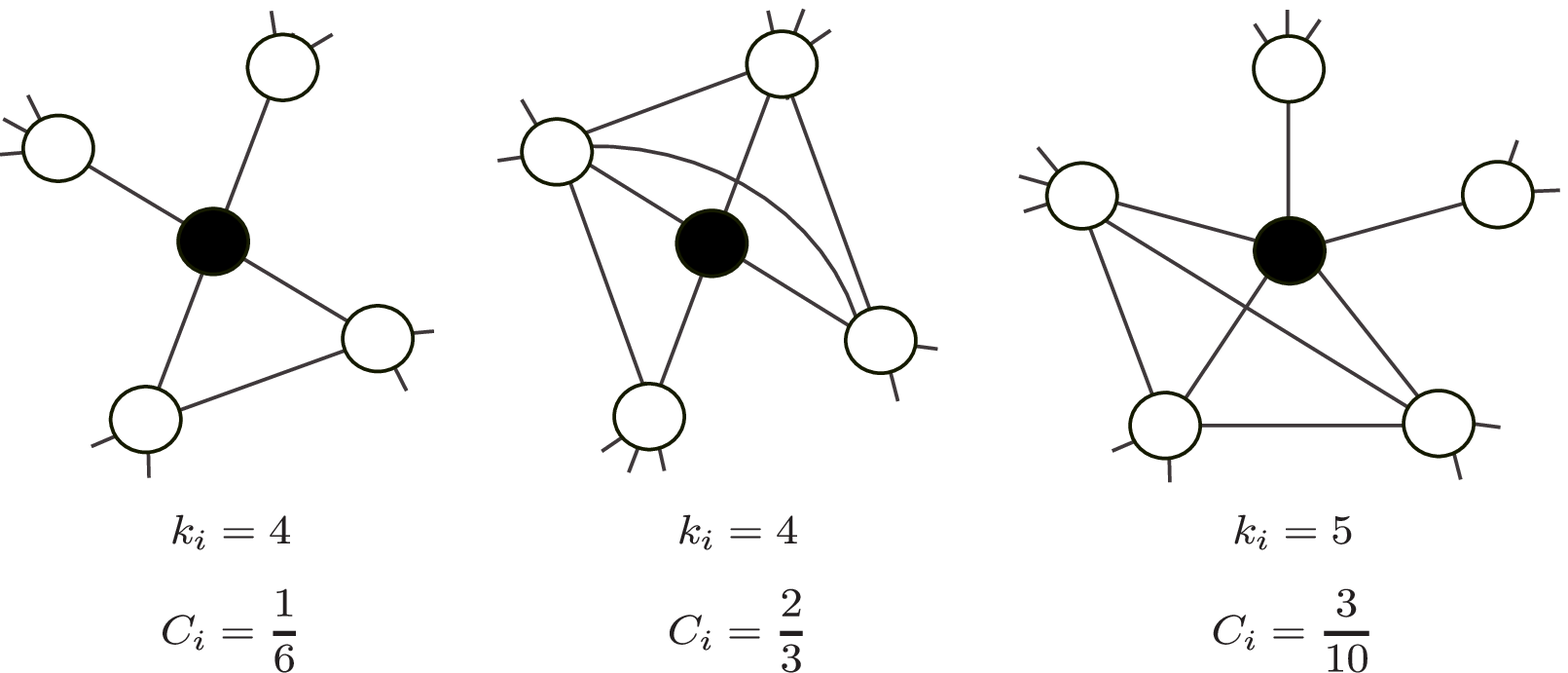}
\end{center}

\clearpage

\begin{flushleft}
Figure 4
\end{flushleft}
\begin{center}
\includegraphics[width=10cm]{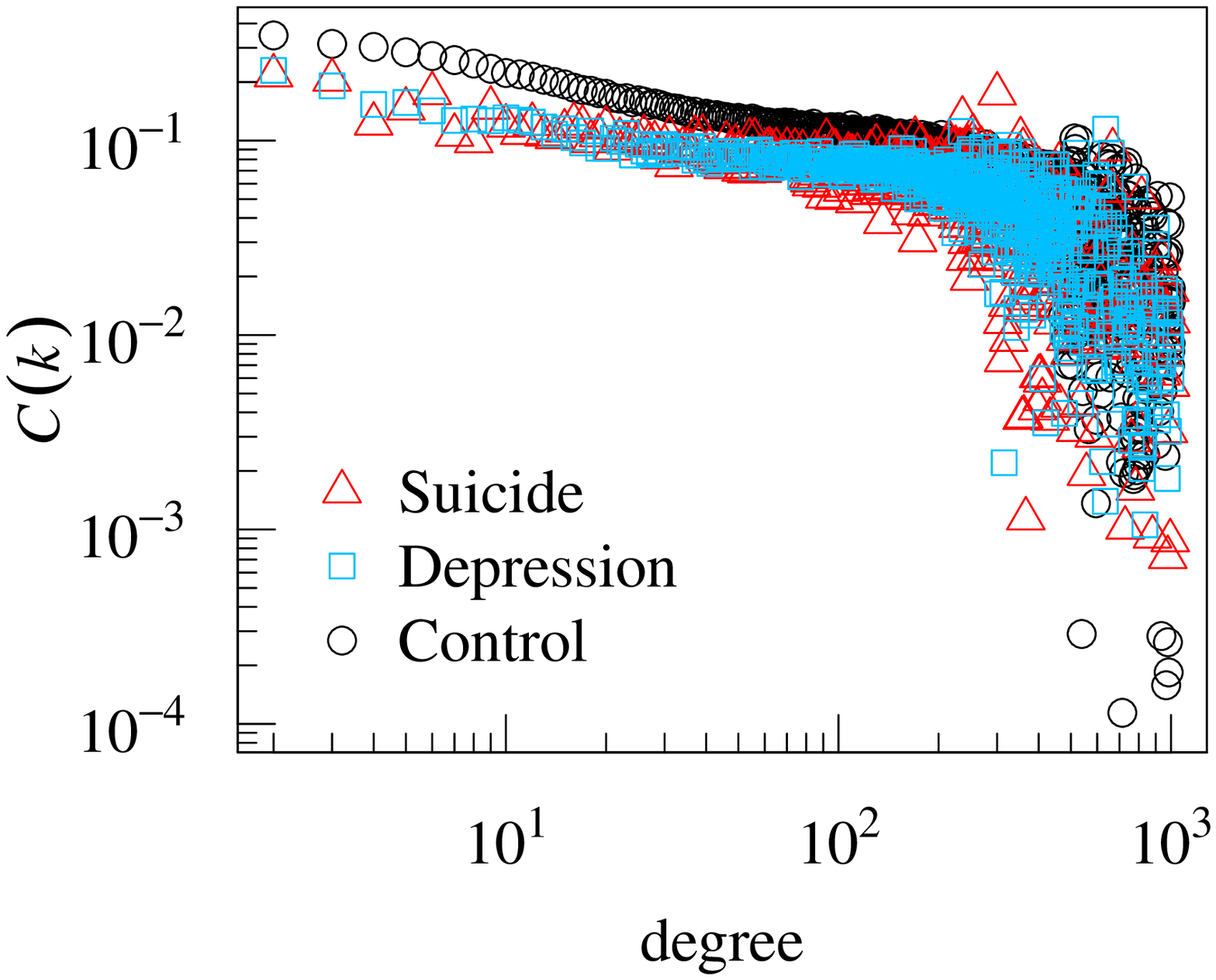}
\end{center}

\clearpage

\newpage

\newlength{\myheight} 

\begin{flushleft}
Table 1
\end{flushleft}

\renewcommand{\arraystretch}{1.2} 
\setlength{\myheight}{1cm} 
\hspace{-3.0cm}
\begin{tabular}{|c|c|c|c|c|c|}
\hline
\raisebox{-3.5ex}[0cm][0cm]{Variable} & 
\multicolumn{2}{c|}{\parbox[c][\myheight][c]{0cm}{} \raisebox{-1.5ex}[0cm][0cm]{\shortstack{Suicide group\\ $ (\textit{N}=9,990) $}}} & 
\multicolumn{2}{c|}{\raisebox{-1.5ex}[0cm][0cm]{\shortstack{Control group\\ $ (\textit{N}=228,949) $}}} & 
\\
\cline{2-5} &
\multicolumn{1}{c|}{\raisebox{-0.5ex}[0cm][0cm]{Mean$\pm$SD}} & 
\multicolumn{1}{c|}{\parbox[c][\myheight][c]{0cm}{} \raisebox{-1.5ex}[0cm][0cm]{\shortstack{Range\\(min,max)}}} &
\multicolumn{1}{c|}{\raisebox{-0.5ex}[0cm][0cm]{Mean$\pm$SD}} & 
\multicolumn{1}{c|}{\parbox[c][\myheight][c]{0cm}{} \raisebox{-1.5ex}[0cm][0cm]{\shortstack{Range\\(min,max)}}} & 
\multicolumn{1}{c|}{\raisebox{3ex}[0cm][0cm]{\textit{p}-value}}
\\ \hline
Age & 27.4$\pm$10.3 & (17, 97) & 27.7$\pm$9.2 & (14, 96) & 0.000652
\\ \hline
Community number & 283.7$\pm$284.3 & (1, 1000) & 46.3$\pm$79.4 & (1, 1000) & $<$ 0.0001
\\ \hline
$k_{i}$ & 82.9$\pm$98.7 & (2, 1000) & 65.8$\pm$67.6 & (2, 1000) & $<$ 0.0001
\\ \hline
$C_{i}$ & 0.087$\pm$0.097 & (0, 1) & 0.150$\pm$0.138 & (0, 1) & $<$ 0.0001
\\ \hline
Homophily (suicide) & 0.0110$\pm$0.0329 & (0, 1.000) & 0.0012$\pm$0.0080 & (0, 0.667) & $<$ 0.0001
\\ \hline
Registration period & 1235.7$\pm$638.9 & (122, 2878) & 1333.5$\pm$670.5 & (102, 2891) & $<$ 0.0001
\\ \hline\hline
Gender (female) & \multicolumn{2}{c|}{5,786 (57.9$\%$)} & \multicolumn{2}{c|}{126,941 (55.4$\%$)} & $<$ 0.0001
\\ \hline\hline
No. suicidal communities & 1.20$\pm$0.51 & (1, 4) & N/A & N/A & N/A
\\ \hline
No. login days & 28.9$\pm$4.4 & (1, 31) & 26.9$\pm$6.3 & (1, 31) & $<$ 0.0001
\\ \hline
\end{tabular}

\newpage

\begin{flushleft}
Table 2
\end{flushleft}

\renewcommand{\arraystretch}{1.2} 
\begin{tabular}{|c|c|c|c|c|}
\hline
\multicolumn{1}{|c|}{Variable} &
\multicolumn{1}{c|}{OR} & 
\multicolumn{1}{c|}{CI} &  
\multicolumn{1}{c|}{$\textit{p}$-value} & 
\multicolumn{1}{c|}{VIF}
\\ \hline
Age & 1.00463 & (1.00211, 1.00716) & 0.000313 & 1.091
\\ \hline
Gender (female = 1) & 0.821 & (0.783, 0.861) & $<$ 0.0001 & 1.028
\\ \hline
Community number & 1.00733 & (1.00720, 1.00747) & $<$ 0.0001 & 1.197
\\ \hline
$k_{i}$ & 0.99790 & (0.99758, 0.99821) & $<$ 0.0001 & 1.156
\\ \hline
$C_{i}$ & 0.0093 & (0.0069, 0.0126) & $<$ 0.0001 & 1.081
\\ \hline
Homophily (suicide) & $2.22\times10^{12}$ & $(0.57\times10^{12}, 8.65\times10^{12})$ & $<$ 0.0001 & 1.016
\\ \hline
Registration period & 0.999383 & (0.999346, 0.999420) & $<$ 0.0001 & 1.135
\\ \hline
\end{tabular}
\newpage

\begin{flushleft}
Table 3
\end{flushleft}

\hspace{-3.0cm}
\scalebox{0.8}{
\begin{tabular}{|c|c|r|r|r|r|r|r|r|r|r|}
\hline
\raisebox{-1.5ex}[0cm][0cm]{Variable 1} & 
\raisebox{-1.5ex}[0cm][0cm]{Variable 2}&
\multicolumn{3}{c|}{Suicide} & 
\multicolumn{3}{c|}{Depression} & 
\multicolumn{3}{c|}{Control}\\
\cline{3-11} & & 
\multicolumn{1}{c|}{P} & \multicolumn{1}{c|}{S} &\multicolumn{1}{c|}{K} & 
\multicolumn{1}{c|}{P} &\multicolumn{1}{c|}{S} & \multicolumn{1}{c|}{K} &
\multicolumn{1}{c|}{P} & \multicolumn{1}{c|}{S} &\multicolumn{1}{c|}{K}  
\\ \hline
Age & Gender & $ -.094 $ & $ -.137 $ & $ -.116 $ & $ -.166 $ & $ -.174 $ & $ -.145 $ & $ -.053 $ & $ -.026 $ & $ -.022 $
\\ \hline
Age & Community number & $ -.045 $ & $ -.105 $ & $ -.073 $ & $ -.089 $ & $ -.131 $ & $ -.091 $ & $ -.032 $ & $ .023 $ & $ .015 $
\\ \hline
Age & $ k_{i} $ & $ -.103 $ & $ -.224 $ & $ -.157 $ & $ -.168 $ & $ -.268 $ & $ -.187 $ & $ -.279 $ & $ -.385 $ & $ -.271 $
\\ \hline
Age & $ C_{i} $ & $ -.048 $ & $ -.220 $ & $ -.154 $ & $ -.092 $ & $ -.273 $ & $ -.192 $ & $ .041 $ & $ -.152 $ & $ -.111 $
\\ \hline
Age & Homophily (suicide) & $ .031 $ & $ -.037 $ & $ -.029 $ & N/A & N/A & N/A & $ -.011 $ & $ -.090 $ & $ .074 $
\\ \hline
Age & Homophily (depression) & N/A & N/A & N/A & $ .166 $ & $ .121 $ & $ -.089 $ & $ -.007 $ & $ -.083 $ & $ -.066 $
\\ \hline
Age & Registration period & $ .159 $ & $ .356 $ & $ .259 $ & $ .203 $ & $ .364 $ & $ .266 $ & $ .278 $ & $ .460 $ & $ .337 $
\\ \hline
Gender & Community number & $ .205 $ & $ .204 $ & $ .166 $ & $ .086 $ & $ .083 $ & $ .068 $ & $ .110 $ & $ .116 $ & $ .095 $
\\ \hline
Gender & $ k_{i} $ & $ .048 $ & $ .046 $ & $ .038 $ & $ .048 $ & $ .046 $ & $ .038 $ & $ .015 $ & $ .014 $ & $ .011 $
\\ \hline
Gender & $ C_{i} $ & $ -.109 $ & $ -.097 $ & $ -.080 $ & $ -.061 $ & $ -.030 $ & $ -.024 $ & $ -.084 $ & $ -.085 $ & $ -.069 $
\\ \hline
Gender & Homophily (suicide) & $ -.007 $ & $ .031 $ & $ .028 $ & N/A & N/A & N/A & $ -.012 $ & $ -.017 $ & $ -.017 $
\\ \hline
Gender & Homophily (depression) & N/A & N/A & N/A & $ -.053 $ & $ -.021 $ & $ -.018 $ & $ .000 $ & $ .009 $ & $ .008 $
\\ \hline
Gender & Registration period & $ -.064 $ & $ -.061 $ & $ -.050 $ & $ -.078 $ & $ -.079 $ & $ -.065 $ & $ .025 $ & $ .025 $ & $ .020 $
\\ \hline
Community number & $ k_{i} $ & $ .348 $ & $ .338 $ & $ .231 $ & $ .375 $ & $ .360 $ & $ .248 $ & $ .375 $ & $ .372 $ & $ .258 $
\\ \hline
Community number & $ C_{i} $ & $ -.231 $ & $ -.200 $ & $ -.136 $ & $ -.201 $ & $ -.171 $ & $ -.116 $ & $ -.376 $ & $ -.399 $ & $ -.277 $
\\ \hline
Community number & Homophily (suicide) & $ -.034 $ & $ .140 $ & $ .105 $ & N/A & N/A & N/A & $ .027 $ & $ .113 $ & $ .091 $
\\ \hline
Community number & Homophily (depression) & N/A & N/A & N/A & $ -.150 $ & $ .034 $ & $ .025 $ & $ .038 $ & $ .166 $ & $ .132 $
\\ \hline
Community number & Registration period & $ .166 $ & $ .152 $ & $ .102 $ & $ .187 $ & $ .172 $ & $ .115 $ & $ .339 $ & $ .338 $ & $ .230 $
\\ \hline
$ k_{i} $ & $ C_{i} $ & $ -.251 $ & $ -.116 $ & $ -.085 $ & $ -.240 $ & $ -.105 $ & $ -.074 $ & $ -.363 $ & $ -.248 $ & $ -.175 $
\\ \hline
$ k_{i} $ & Homophily (suicide) & $ -.175 $ & $ .174 $ & $ .107 $ & N/A & N/A & N/A & $ -.013 $ & $ .191 $ & $ .150 $
\\ \hline
$ k_{i} $ & Homophily (depression) & N/A & N/A & N/A & $ -.210 $ & $ .076 $ & $ .029 $ & $ -.027 $ & $ .254 $ & $ .188 $
\\ \hline
$ k_{i} $ & Registration period & $ .170 $ & $ .154 $ & $ .103 $ & $ .172 $ & $ .152 $ & $ .101 $ & $ .102 $ & $ .081 $ & $ .055 $
\\ \hline
$ C_{i} $ & Homophily (suicide) & $ -.047 $ & $ -.213 $ & $ -.162 $ & N/A & N/A & N/A & $ -.026 $ & $ -.100 $ & $ -.080 $
\\ \hline
$ C_{i} $ & Homophily (depression) & N/A & N/A & N/A & $ -.055 $ & $ -.243 $ & $ -.182 $ & $ -.031 $ & $ -.145 $ & $ -.114 $
\\ \hline
$ C_{i} $ & Registration period & $ -.143 $ & $ -.112 $ & $ -.162 $ & $ -.133 $ & $ -.099 $ & $ -.068 $ & $ -.221 $ & $ -.249 $ & $ -.168 $
\\ \hline
Homophily (suicide) & Registration period & $ -.104 $ & $ -.059 $ & $ -.044 $ & N/A & N/A & N/A & $ -.039 $ & $ -.031 $ & $ -.025 $
\\ \hline
Homophily (depression) & Registration period & N/A & N/A & N/A & $ -.120 $ & $ -.049 $ & $ -.036 $ & $ -.024 $ & $ .011 $ & $ .009 $
\\ \hline
\end{tabular}
}

\newpage

\begin{flushleft}
Table 4
\end{flushleft}

\renewcommand{\arraystretch}{1.2} 
\begin{tabular}{|c|c|c|c|c|c|}
\hline
\multicolumn{1}{|c|}{Variable} &
\multicolumn{1}{c|}{OR} & 
\multicolumn{1}{c|}{CI} &  
\multicolumn{1}{c|}{$\textit{p}$-value} & 
\multicolumn{1}{c|}{AUC}
\\ \hline
Age & 0.99604 & (0.99377, 0.99832) & 0.000651 & 0.515
\\ \hline
Gender (female = 1) & 1.106 & (1.062, 1.152) & $<$ 0.0001 & 0.512
\\ \hline
Community number & 1.00728 & (1.00716, 1.00741) & $<$ 0.0001 & 0.867
\\ \hline
$k_{i}$ & 1.00259 & (1.00237, 1.00280) & $<$ 0.0001 & 0.549
\\ \hline
$C_{i}$ & 0.000581 & (0.000428, 0.000789) & $<$ 0.0001 & 0.690
\\ \hline
Homophily (suicide) & $1.57\times10^{16}$ & $(0.41\times10^{16}, 6.08\times10^{16})$ & $<$ 0.0001 & 0.643
\\ \hline
Registration period & 0.999783 & (0.999753, 0.999813) & $<$ 0.0001 & 0.545
\\ \hline
\end{tabular}

\newpage

\begin{flushleft}
Table 5
\end{flushleft}

\renewcommand{\arraystretch}{1.2}
\setlength{\myheight}{1cm}
\hspace{-3.0cm}
\begin{tabular}{|c|c|c|c|c|c|}
\hline
\raisebox{-3.5ex}[0cm][0cm]{Variable} & 
\multicolumn{2}{c|}{\parbox[c][\myheight][c]{0cm}{} \raisebox{-1.5ex}[0cm][0cm]{\shortstack{Depression group\\ $ (\textit{N}=24,410) $}}} & 
\multicolumn{2}{c|}{\raisebox{-1.5ex}[0cm][0cm]{\shortstack{Control group\\ $ (\textit{N}=228,949) $}}} & 
\\
\cline{2-5} &
\multicolumn{1}{c|}{\raisebox{-0.5ex}[0cm][0cm]{Mean$\pm$SD}} & 
\multicolumn{1}{c|}{\parbox[c][\myheight][c]{0cm}{} \raisebox{-1.5ex}[0cm][0cm]{\shortstack{Range\\(min,max)}}} &
\multicolumn{1}{c|}{\raisebox{-0.5ex}[0cm][0cm]{Mean$\pm$SD}} & 
\multicolumn{1}{c|}{\parbox[c][\myheight][c]{0cm}{} \raisebox{-1.5ex}[0cm][0cm]{\shortstack{Range\\(min,max)}}} & 
\multicolumn{1}{c|}{\raisebox{3ex}[0cm][0cm]{\textit{p}-value}}
\\ \hline
Age & 28.8$\pm$9.4 & (16, 97) & 27.7$\pm$9.2 & (14, 96) & $<$ 0.0001
\\ \hline
Community number & 249.6$\pm$263.1 & (1, 1000) & 46.3$\pm$79.4 & (1, 1000) & $<$ 0.0001
\\ \hline
$k_{i}$ & 81.9$\pm$88.1 & (2, 1000) & 65.8$\pm$67.6 & (2, 1000) & $<$ 0.0001
\\ \hline
$C_{i}$ & 0.085$\pm$0.089 & (0, 1) & 0.150$\pm$0.138 & (0, 1) & $<$ 0.0001
\\ \hline
Homophily (depression) & 0.0196$\pm$0.0501 & (0, 1.000) & 0.0031$\pm$0.0131 & (0, 0.667) & $<$ 0.0001
\\ \hline
Registration period & 1389.4$\pm$659.2 & (122, 2885) & 1333.5$\pm$670.5 & (102, 2891) & $<$ 0.0001
\\ \hline\hline
Gender (female) & \multicolumn{2}{c|}{16,872 (69.1$\%$)} & \multicolumn{2}{c|}{126,941 (55.4$\%$)} & $<$ 0.0001
\\ \hline\hline
No. suicidal communities  & 1.16$\pm$0.47 & (1, 6) & N/A & N/A & N/A
\\ \hline
No. login days & 28.8$\pm$4.4 & (1, 31) & 26.9$\pm$6.3 & (1, 31) & $<$ 0.0001
\\ \hline
\end{tabular}

\newpage

\begin{flushleft}
Table 6
\end{flushleft}

\renewcommand{\arraystretch}{1.2}
\begin{tabular}{|c|c|c|c|c|}
\hline
\multicolumn{1}{|c|}{Variable} &
\multicolumn{1}{c|}{OR} & 
\multicolumn{1}{c|}{CI} &  
\multicolumn{1}{c|}{$\textit{p}$-value} & 
\multicolumn{1}{c|}{VIF}
\\ \hline
Age & 1.0141 & (1.0124, 1.0158) & $<$ 0.0001 & 1.104
\\ \hline
Gender (female = 1) & 1.532 & (1.481, 1.585) & $<$ 0.0001 & 1.019
\\ \hline
Community number & 1.00790 & (1.00778, 1.00803) & $<$ 0.0001 & 1.155
\\ \hline
$k_{i}$ & 0.99833 & (0.99810, 0.99856) & $<$ 0.0001 & 1.154
\\ \hline
$C_{i}$ & 0.0145 & (0.0118, 0.0178) & $<$ 0.0001 & 1.079
\\ \hline
Homophily (depression) & $1.98\times10^{10}$ & $(0.99\times10^{10}, 4.02\times10^{10})$ & $<$ 0.0001 & 1.022
\\ \hline
Registration period & 0.999744 & (0.999720, 0.999769) & $<$ 0.0001 & 1.117
\\ \hline
\end{tabular}

\newpage

\begin{flushleft}
Table 7
\end{flushleft}

\renewcommand{\arraystretch}{1.2}
\begin{tabular}{|c|c|c|c|c|c|}
\hline
\multicolumn{1}{|c|}{Variable} &
\multicolumn{1}{c|}{OR} & 
\multicolumn{1}{c|}{CI} &  
\multicolumn{1}{c|}{$\textit{p}$-value} & 
\multicolumn{1}{c|}{AUC}
\\ \hline
Age & 1.0110 & (1.0097, 1.0123) & $<$ 0.0001 & 0.551
\\ \hline
Gender (female = 1) & 1.799 & (1.748, 1.850) & $<$ 0.0001 & 0.568
\\ \hline
Community number & 1.00826 & (1.00814, 1.00837) & $<$ 0.0001 & 0.860
\\ \hline
$k_{i}$ & 1.00258 & (1.00243, 1.00274) & $<$ 0.0001 & 0.566
\\ \hline
$C_{i}$ & 0.000415 & (0.000338, 0.000509) & $<$ 0.0001 & 0.692
\\ \hline
Homophily (depression) & $2.12\times10^{12}$ & $(1.05\times10^{12}, 4.28\times10^{12})$ & $<$ 0.0001 & 0.658
\\ \hline
Registration period & 1.000126 & (1.000106, 1.000145) & $<$ 0.0001 & 0.522
\\ \hline
\end{tabular}

\newpage

\begin{flushleft}
Table 8
\end{flushleft}

\setlength{\myheight}{1cm} 
\hspace{-3.0cm}
\begin{tabular}{|c|c|c|c|c|c|c|}
\hline
\multirow{2}{*}{ID} & Date of creation &
\multirow{2}{*}{No. users} &
No. active & 
Fraction of & No. & No. active \\[-0.12cm]
& (day/month/year) && users & active users ($\%$) & comments & topics
\\ \hline
1 & 18/01/2008& 8367 & 5985 & 69.9 & 741 & 16
\\ \hline
2 & 21/09/2006 & 5135 & 3192 & 62.9 & 318 & 6
\\ \hline
3 & 01/12/2004 & 3459 & 1883 & 53.2 & 279 & 12
\\ \hline
4 & 04/02/2008 & 1445 & 965 & 62.4 & 105 & 9
\\ \hline
\end{tabular}

\newpage

\begin{flushleft}
Table 9
\end{flushleft}

\setlength{\myheight}{1cm}
\hspace{-3.0cm}
\begin{tabular}{|c|c|c|c|c|c|c|}
\hline
\multirow{2}{*}{ID} & Date of creation &
\multirow{2}{*}{No. users} &
No. active & 
Fraction of & No. & No. active \\[-0.12cm]
& (day/month/year) && users & active users ($\%$) & comments & topics
\\ \hline
1 & 06/04/2004 & 15618 & 8605 & 54.7 & 14466 & 52
\\ \hline
2 & 06/02/2006 & 13082 & 9674 & 72.8 & 1008 & 16
\\ \hline
3 & 08/12/2004 & 4948 & 2845 & 56.5 & 782 & 17
\\ \hline
4 & 22/04/2006 & 4606 & 2907 & 60.4 & 221 & 30
\\ \hline
5 & 28/01/2008 & 3406 & 2321 & 65.0 & 1350 & 24
\\ \hline
6 & 09/12/2004 & 3464 & 2039 & 58.2 & 851 & 20
\\ \hline
7 & 21/12/2004 & 2440 & 1367 & 54.2 & 535 & 5
\\ \hline
\end{tabular}

\end{document}